\documentclass[]{article}
\usepackage[utf8]{inputenc}
\usepackage{setspace}
\usepackage{epsfig}
\usepackage{amsfonts}
\usepackage{amssymb}
\usepackage{amsmath}
\usepackage{subfigure}
\usepackage[usenames]{color}
\usepackage{comment}
\usepackage{multirow}
\usepackage{psfrag}
\usepackage{fullpage}
\usepackage{verbatim} 
\usepackage{graphicx,epsfig}
\usepackage{fixltx2e}
\usepackage{units}
\usepackage[centerlast]{caption}
\usepackage{bytefield}
\usepackage{pgf}
\usepackage{tikz}
\usetikzlibrary{arrows,automata}
\usepackage{soul}
\usepackage{pifont}
\usepackage{slashbox}
\usepackage{url}
\usepackage{verbatim}

\newcommand{\UL}{\text{UL}}
\newcommand{\DL}{\text{DL}}

\DeclareGraphicsExtensions{.eps,.jpg,.eps}

\begin{document}

\title{IEEE 802.11ah: The Wi-Fi Approach\\ for M2M Communications}
\author{T. Adame, A. Bel, B. Bellalta, J. Barcelo, M. Oliver\\NeTS Research Group \\Universitat Pompeu Fabra, Barcelona\\
	    \emph{\{toni.adame, albert.bel, boris.bellalta, jaume.barcelo, miquel.oliver\}@upf.edu}}

\maketitle

\begin{abstract}

Machine-to-Machine (M2M) communications are positioned to be one of the fastest growing technology segments in the next decade. Sensor and actuator networks connect communication machines and devices so that they automatically transmit information, serving the growing demand for environmental data acquisition. The IEEE 802.11ah Task Group (TGah) is working on a new standard to address the particular requirements of M2M networks: a large number of power-constrained stations, a long transmission range, small and infrequent data messages, low data rates and a non-critical delay. This paper explores the key features of this new standard, especially those related to the reduction of energy consumption in the medium access control layer. Given these requirements, a performance assessment of IEEE 802.11ah in four common M2M scenarios such as agriculture monitoring, smart metering, industrial automation and animal monitoring is presented.

\end{abstract}


\doublespace

\section{Introduction}

Several studies have forecasted an annual growth rate over 20\% in the number of M2M (Machine-to-Machine) connections globally \cite{hatton}, with more than 10 billion mobile-connected devices, exceeding the world’s population, in 2017.

The communication technologies currently used for M2M applications can be classified in two categories:  Wireless Sensor Networks (WSNs), for interconnecting multiple sensor nodes spread over a particular area; and regular mobile (cellular) networks, for isolated/scattered nodes or to allow the gateway of a particular WSN to reach the Internet \cite{Babamir12}. 

\begin{figure}[h!]
\centering
\includegraphics[width=16cm]{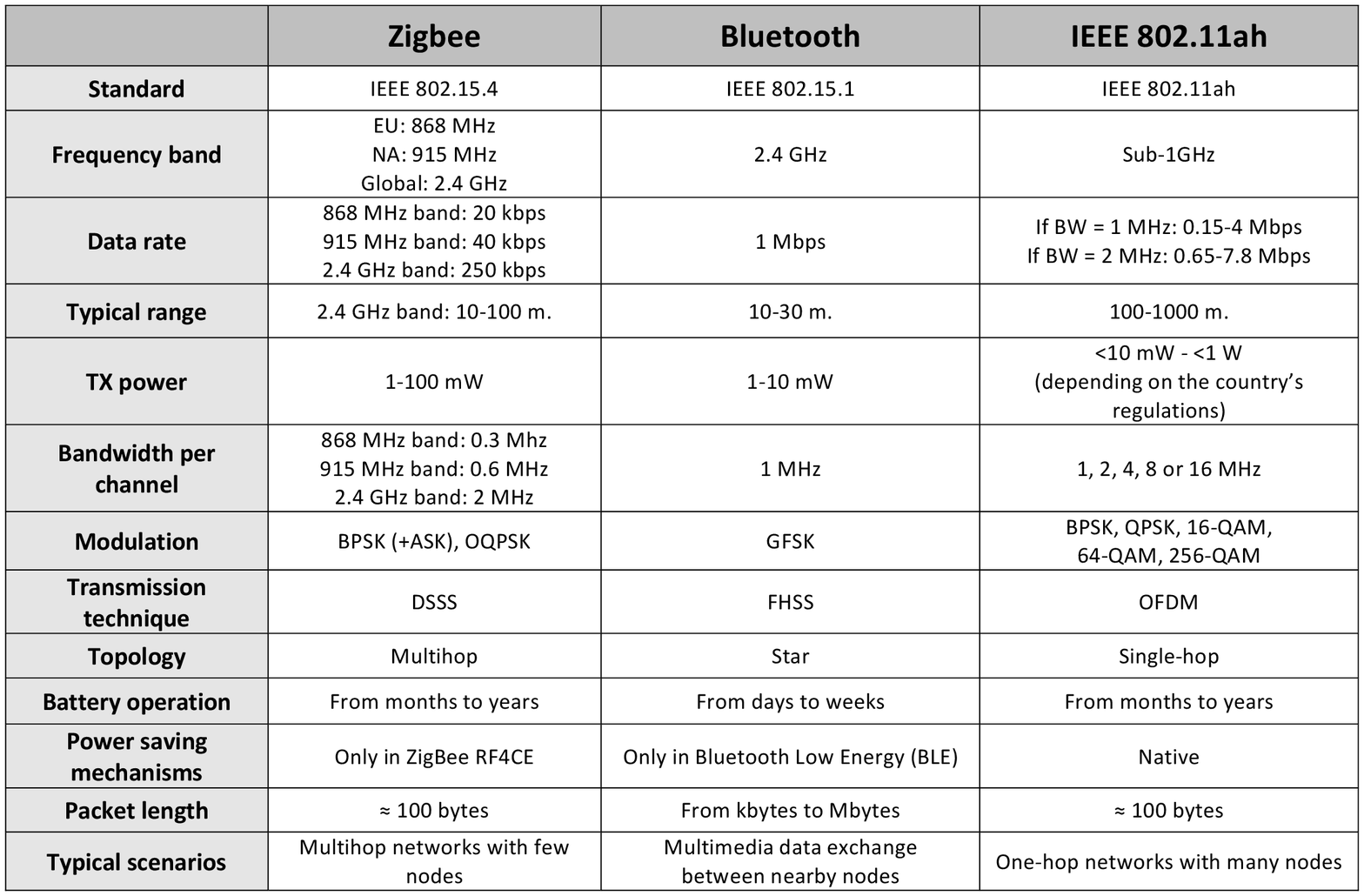}
\caption{Comparison of different unlicensed M2M technologies (based on \cite{badihy})}
\label{fig:comp_table}
\end{figure}

With respect to WSNs, different systems (Zigbee, 802.15.4, 6LoWPAN, Bluetooth or even proprietary radio solutions) have been considered for transmitting data in common M2M scenarios (see Table \ref{fig:comp_table}). However, none of those systems has prevailed because of the current diversity and complexity of the applications and environments. As for mobile networks, M2M communications are currently mainly supported by GPRS/EDGE networks because of the growing but still reduced number of devices and light traffic requirements. Simultaneously, the 3rd Generation Partnership Project (3GPP) is working towards supporting M2M applications on 4G broadband mobile networks, such as UMTS and LTE, with the goal of natively embedding M2M communications in the upcoming 5G systems.
\vspace{2cm}
\renewcommand{\figurename}{Table}%
\renewcommand{\figurename}{Figure}%

The IEEE 802.11ah Task Group (TGah), created in 2010, addresses the need for an M2M wireless standard to cover the existing gap between traditional mobile networks and the growing demand for wireless sensor networks. TGah deals with the specification of an unlicensed sub-1GHz worldwide wireless local area network (WLAN) standard for future M2M communications supporting a wide set of scenarios based on a large number of devices, a long range and energy constraints. IEEE 802.11ah offers a simple, robust and efficient solution in the ISM band compared with existing WSNs. Moreover, the IEEE 802.11ah specification points out a comparable high quality of service to the one provisioned by current mobile networks, building a completely scalable and cost-effective operation. 

This paper introduces the IEEE 802.11ah \cite{aust2012ieee,aust2012sub} amendment, which allows WLANs to manage hundreds or even thousands of low-capability M2M devices with sporadic traffic needs. Although IEEE 802.11 has become the dominant standard for WLANs and one of the most commonly deployed technologies, we expect the new release of IEEE 802.11ah to be vastly adopted as a legacy solution to deploy WSNs in most markets and applications.

The goal of this paper, apart from introducing the main features of the IEEE 802.11ah amendment, is to analyze its feasibility and evaluate its performance in four central M2M data acquisition application areas: agriculture monitoring, smart metering, industrial automation and animal monitoring.

The remainder of this paper is organized as follows. In Section \ref{design}, IEEE 802.11ah general scenarios and requirements are introduced. Section \ref{technology} describes the main features of the amendment in terms of the PHY and MAC layers. In Section \ref{proposal}, we introduce the different scenarios used in our evaluation as well as the results obtained for different figures of merit. Finally, Section \ref{conclusions} presents the conclusions and discusses open challenges.


\section{Scenarios and Requirements} \label{design}

M2M communications include any technology that enables devices to exchange information and perform actions without human intervention. It is expected that M2M communications will be one of the major technological drivers in the next decade, mainly in the following areas: metering and control of utilities (electricity, gas, heat, and water), home and industrial automation, eHealth, surveillance, and intelligent transport systems.

TGah has defined several application areas that would motivate the use of this new sub-1GHz standard \cite{aust2012ieee}:

\begin{enumerate}
\item Sensor networks
\par Current Wi-Fi networks cannot sufficiently support sensor networks for three main reasons:

\begin{itemize}
\item \underline{The absence of power saving mechanisms:} The particular energy constraints of sensor networks are not taken into account in the IEEE 802.11 standard; the standard does not include energy saving mechanisms specially designed for these types of devices.

\item \underline{The use of unsuitable bands:} Due to their short wireless range and high obstruction losses, current Wi-Fi bands require the use of intermediate nodes, adding complexity to the network. Implicitly, this puts the focus on the lack of an IEEE 802.11 standardized implementation in a band more suitable for low-rate and long-range networks.

\item \underline{The existence of low cost alternatives:} The limited use of Wi-Fi for data communication between low-capability and battery-powered nodes has led to an upsurge in low power alternatives, such as IEEE 802.15.4, 6LoWPAN, Zigbee, and sub-1GHz proprietary protocols, which are all categorized as WSNs.
\end{itemize}

\item Backhaul networks for sensors
\par Not only as the final infrastructure but also as backhaul, IEEE 802.11ah networks could exploit their large coverage and act as an intermediate step in the communication between devices (IEEE 802.15.4 nodes, for example) and data collectors.
\end{enumerate}

The requirements defined by IEEE 802.11ah to support M2M communications are as follows \cite{aust2012ieee,aust2012sub}:

\begin{itemize}
\item Up to 8,191 devices associated with an access point (AP) through a hierarchical identifier structure.
\item Carrier frequencies of approximately 900 MHz (license-exempt) that are less congested and guarantee a long range.
\item Transmission range up to 1 km in outdoor areas.
\item Data rates of at least 100 kbps.
\item One-hop network topologies.
\item Short and infrequent data transmissions (data packet size approximately 100 bytes and packet inter-arrival time greater than $30$ s.).
\item Very low energy consumption by adopting power saving strategies.
\item Cost-effective solution for network device manufacturers.
\end{itemize}


\section{Main Technological Features} \label{technology}

To satisfy the requirements defined in the previous section, IEEE 802.11ah designs new PHY and MAC layers. These new layers include several modifications with respect to consolidated IEEE standards -particularly at the MAC level- for supporting the special constraints of M2M communications. 

The IEEE 802.11ah PHY layer can be considered a sub-1GHz version of the PHY layer on the IEEE 802.11ac. Similarly, the IEEE 802.11ah MAC layer incorporates most of the main IEEE 802.11 characteristics, adding some novel power management mechanisms.

\subsection{PHY Layer}

IEEE 802.11ah operates over a set of unlicensed radio bands (all sub-1GHz) that depend on country regulations. For example, the targeted frequency bands are 863-868 MHz in Europe, 902-928 MHz in the US and 916.5-927.5 MHz in Japan. China, South Korea and Singapore also have specific allocations \cite{aust2012ieee}. Channel bandwidths of 1 MHz and 2 MHz have been widely adopted, although in some countries broader configurations using 4, 8 and 16 MHz are also allowed. 

PHY transmission is an OFDM-based waveform consisting of 32 or 64 tones/sub-carriers (including tones allocated as pilots, guard and Direct Current) with 31.25 kHz spacing. The supported modulations include BPSK, QPSK and from 16 to 256-QAM. Technologies such as single-user beamforming, Multi Input Multi Output (MIMO) and downlink multi-user MIMO -first introduced in IEEE 802.11ac- are also adopted within the IEEE 802.11ah standard.

\subsection{MAC Layer}
\label{MAC}

The MAC Layer is designed to maximize the number of stations supported by the network while ensuring minimum energy consumption. IEEE 802.11ah defines three different types of stations \cite{Draft802.11ah}, each with different procedures and time periods to access the common channel (see Figure \ref{fig:todo}): Traffic indication map (TIM) stations, non-TIM stations, and unscheduled stations.

\begin{itemize}

\item {Traffic indication map (TIM) stations}
\par This is the only type of station that needs to listen to AP beacons to send or receive data. Their data transmissions must be performed within a restricted access window (RAW) period with three differentiated segments (multicast, downlink and uplink). 
Stations with a high traffic load should use this procedure to access the channel because it combines periodic data transmission segments with energy efficiency mechanisms. These novel features are described in detail below.


\item {Non-TIM stations}
\par Non-TIM stations do not need to listen to any beacons to transmit data. During the association process, non-TIM devices directly negotiate with the AP to obtain a transmission time allocated in a periodic restricted access window (PRAW). The following transmissions can be either periodically defined or renegotiated, depending on the requirements set by the station. Although non-TIM stations can transmit data periodically, it is advisable to deploy TIM stations for high-volume data applications to achieve better management of channel resources and benefit from all the improvements developed by IEEE 802.11ah.

\item {Unscheduled stations}
\par These stations do not need to listen to any beacons, similar to non-TIM stations. Even inside any restricted access window, they can send a poll frame to the AP asking for immediate access to the channel. The response frame indicates an interval (outside both restricted access windows) during which unscheduled stations can access the channel. This procedure is meant for stations that want to sporadically join the network.

\end{itemize}

\subsubsection{Support of Many Associated TIM stations}

One of the biggest challenges for the adoption of legacy IEEE 802.11 is the low number of stations that can be simultaneously associated with the same AP. Henceforth, TGah includes a novel hierarchical method that defines groups of stations and allows to support a larger number of devices. Each group of devices may be configured depending on the type of application used, the power level required or even the targeted QoS (Figure \ref{fig:aid_map}). TIM stations have the following two main characteristics \cite{Draft802.11ah}:

\begin{enumerate}
\item The design of a new association identifier (AID) that classifies stations into pages, blocks (hereafter called \textit{TIM groups}), sub-blocks and stations' indexes in sub-blocks (see Figure  \ref{fig:aid_structure}).
\item The division of the partial virtual bitmap belonging to the TIM information element into smaller bitmaps, one for each TIM group.
\end{enumerate}

During the association stage, the AP allocates an AID to each station; the AID structure is detailed in Figure \ref{fig:aid_structure}. This AID is unique for each station and consists of 13 bits that include the different hierarchical levels, so that the maximum number of supported stations is increased more than 4-fold from the 2,007 of IEEE 802.11 to 8,191 ($= 2^{13}-1$) in IEEE 802.11ah.

\setcounter{figure}{0}
\begin{figure}[p]
\begin{center}
\subfigure[Distribution of channel access restricted windows (RAW and PRAW) among signaling beacons\label{fig:todo}]
{\includegraphics[width=0.85\textwidth]{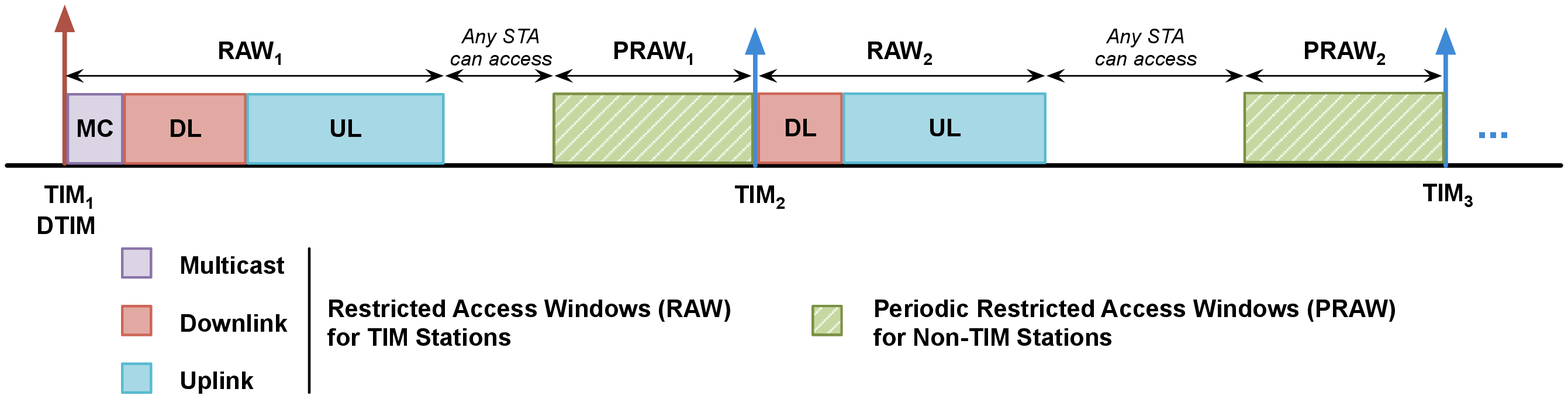}}
\subfigure[AID frame structure based on the hierarchical association of stations \label{fig:aid_structure}]
{\includegraphics[width=0.5\textwidth]{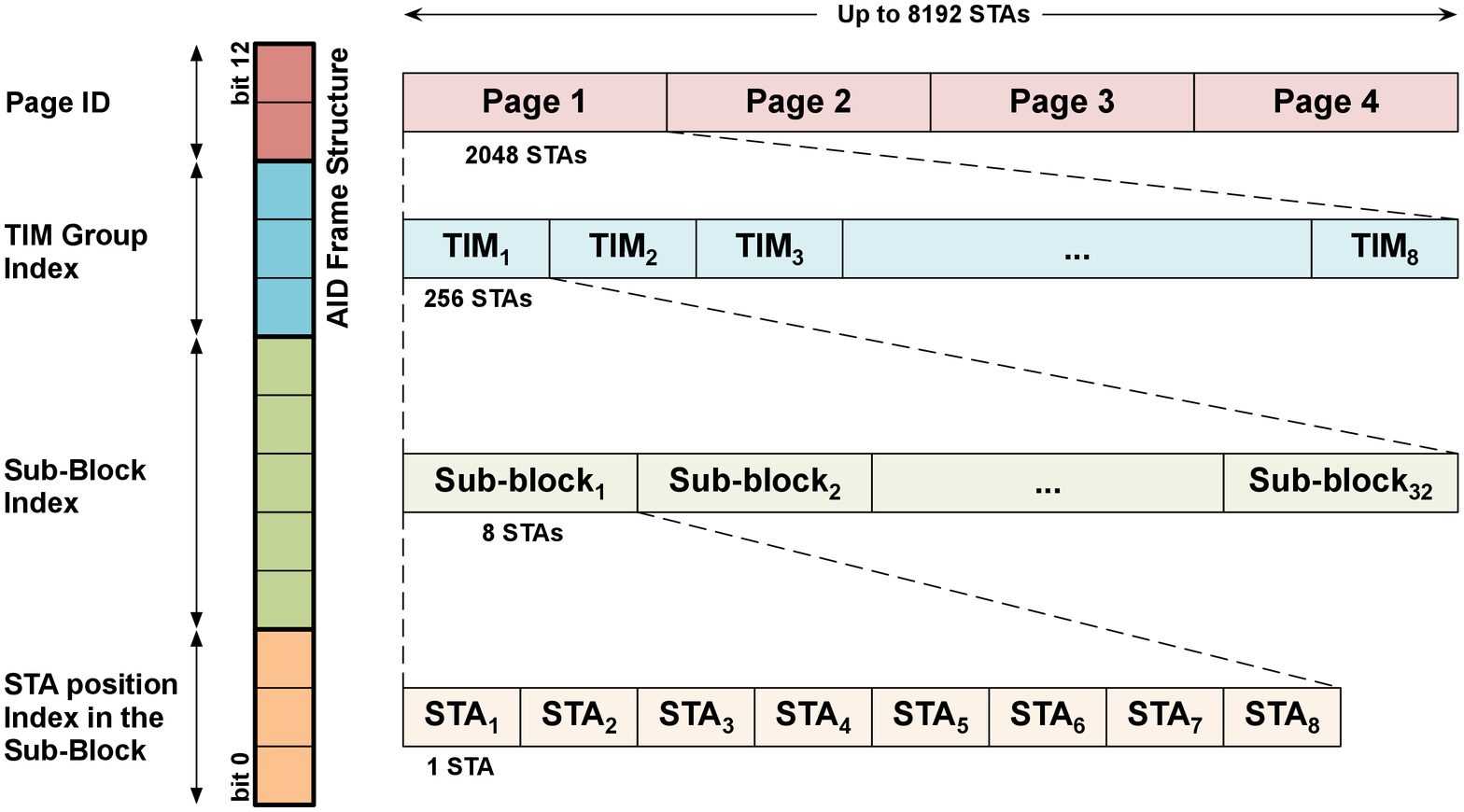}}
\subfigure[AID stations map \label{fig:aid_map}]
{\includegraphics[width=0.4\textwidth]{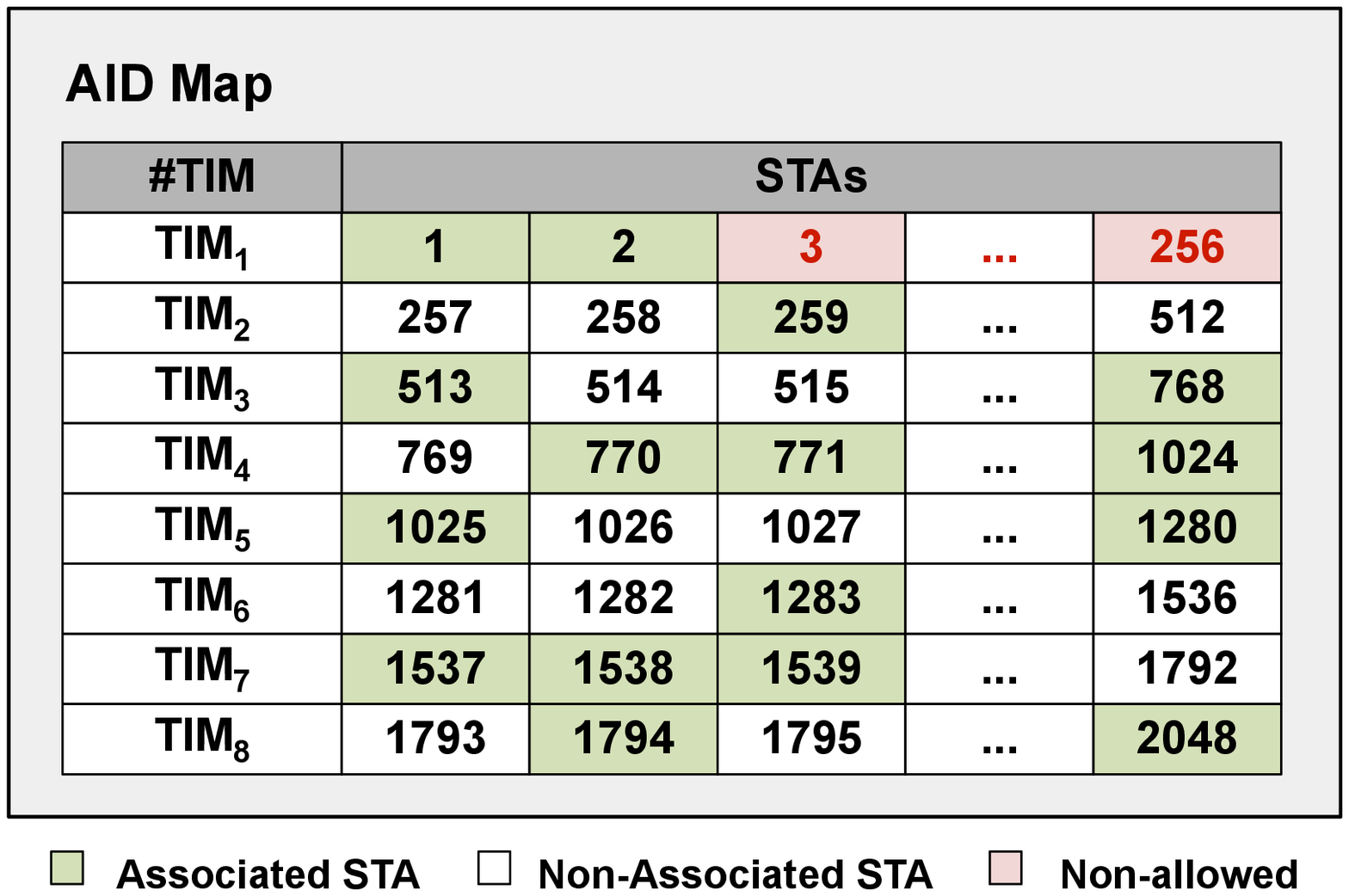}}
\centering\subfigure[Effect of DTIM and TIM mapping over stations. The \textit{TIM$_{7}$ group is signaled in DTIM map, so that all stations in that group listen to TIM$_{7}$ beacon. Only stations \#1538 and \#1539 are signaled in TIM$_{7}$ beacon and, therefore have a contention time for accessing the channel.} \label{fig:DTIM_effects}]
{\includegraphics[scale=0.7]{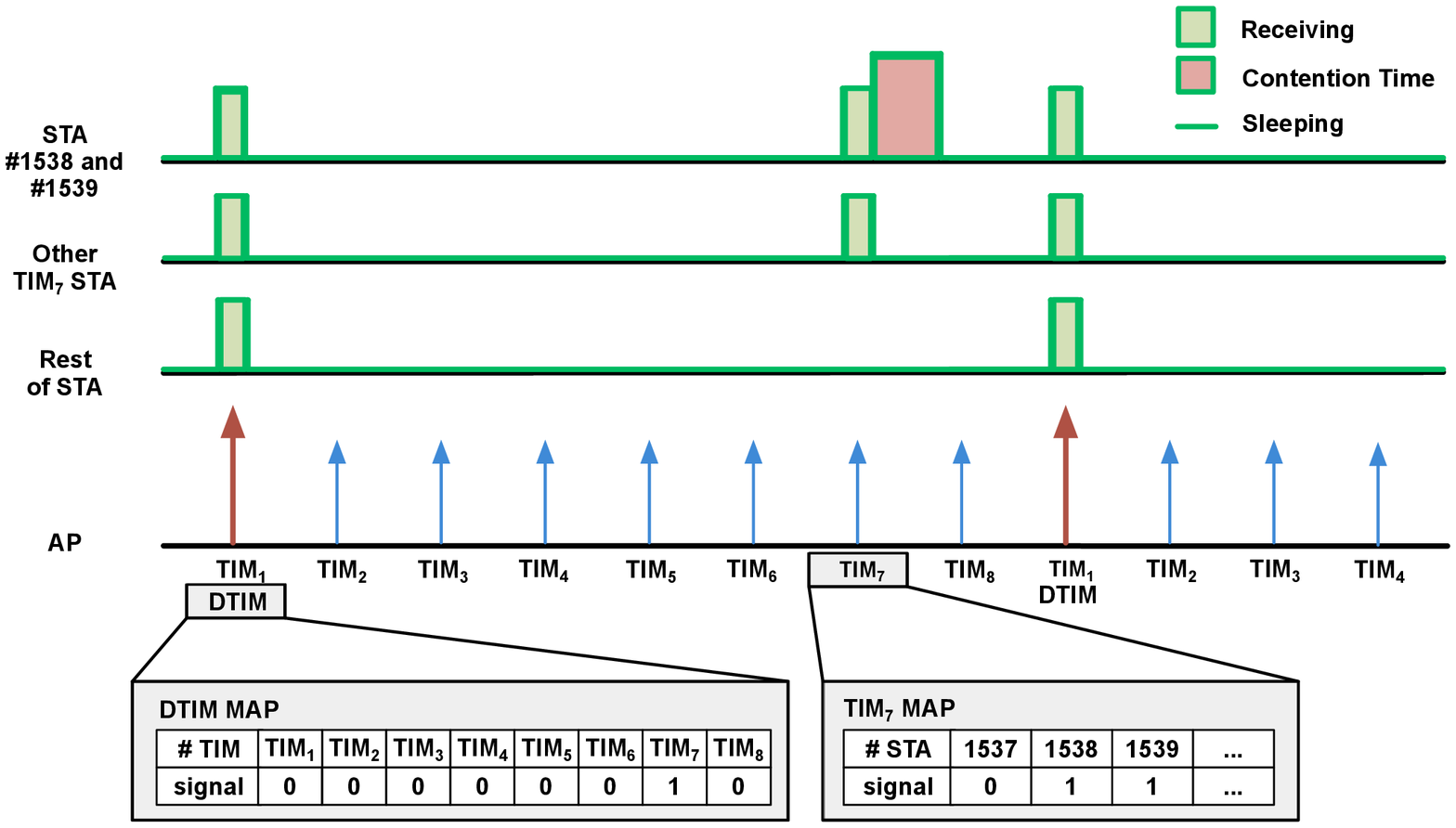}}
\caption{IEEE 802.11ah new MAC contributions regarding to distribution of stations and signaling} \label{fig:Figure1}
\end{center}
\end{figure}

\subsubsection{Power Saving Mechanisms for TIM stations}
IEEE 802.11ah includes a power saving mode to reduce the energy consumption of sensor devices. This mechanism exploits the low-power states of the network interface by deactivating the radio module during non-traffic periods.

\paragraph{TIM and Page Segmentation} 

To reduce the time a station is competing for the channel as well as to increase its sleep time, IEEE 802.11ah uses a scheme called \textit{TIM and Page Segmentation}. Thus, the hierarchical distribution of stations into groups is used not only for organizational purposes but also for scheduling signaling and allocating available channel resources to the different TIM groups.

IEEE 802.11ah restricts to a particular TIM group of stations to, simultaneously, contend for the same channel in a specific period. These stations only need to compete among themselves and listen to their associated TIM beacons. This access mechanism allows TIM stations from the same group to be in a sleep mode for the rest of the time, resulting in a significant reduction of the energy consumption.

The signaling system defined is an extension of the IEEE 802.11 one; in addition to using TIM beacons for \textit{station-level signaling}, it also uses delivery traffic indication map (DTIM) beacons for \textit{TIM group-level signaling} (see Figure \ref{fig:DTIM_effects}). The purposes of these two types of beacons are as follows:

\begin{enumerate}
\item \textit{DTIM beacons} are used to signal which TIM groups have pending data, unicast and/or multicast, in the AP. The beacon also contains all information about the restricted access window properties, such as segment durations, sub-slotting mechanisms, etc.
\item \textit{TIM beacons} page a single TIM group with at least one station having pending data in the AP. Between two consecutive DTIM beacons, there are as many TIM beacons as groups defined.
\end{enumerate}

Using this mechanism, any station can enter a power-saving state during the entire restricted access window period if it does not have a packet to transmit and at least one of the following two conditions is met: 1) it observes in the DTIM beacon that there is no downlink traffic addressed to its TIM group or 2) it observes in the DTIM beacon that there is downlink traffic addressed to its TIM group but does not itself explicitly appear in its TIM beacon.

\paragraph{Advanced Signaling Modes}
\label{signaling}
Within the \textit{TIM and Page Segmentation} scheme, when the number of network pages is greater than one, IEEE 802.11ah offers two advanced signaling modes: 
\begin{enumerate}
\item \textit{Non-TIM offset}: The signaling information of a particular TIM group is transmitted in the same beacon as many times as the number of network pages. This is the default mode in IEEE 802.11ah.
\item \textit{TIM offset}: This mode includes a 5-bit field in the DTIM beacon that allows TIM groups from different pages to be separately scheduled over their own TIM beacons.
\end{enumerate}

The TIM offset mode has lower energy consumption than the Non-TIM mode. However, its behavior with respect to the maximum number of stations supported, packet delivery ratio (PDR), and network efficiency is slightly worse  \cite{adame2013}.

\subsubsection{Channel Access for TIM stations}

The IEEE 802.11ah channel access for TIM stations combines an AP-centralized time period allocation system with the distributed coordination function (DCF) medium-access technique within those periods. 

Regardless of the signaling mode employed, the time between consecutive TIMs contains a restricted access window formed by one downlink segment, one uplink segment and one multicast segment placed immediately after each DTIM beacon (in Figure \ref{fig:todo}). 

The data transmission procedures for both the downlink and uplink cases are detailed below:
\begin{itemize}
\item \underline{Downlink:}
When an AP has a packet to be sent to a station, the DTIM beacon must include in its bitmap the TIM group to which that station belongs. Then, the corresponding TIM beacon includes that station in its bitmap. Each signaled station must listen to its TIM beacon to know when to start contending. This contention will be performed using the DCF. When the backoff of a station expires, it sends a PS-Poll frame to obtain its corresponding data.

\item \underline{Uplink:}
Whenever a station wants to send an uplink message to the AP, it must first listen to its corresponding TIM group to know when to contend for the channel. The contention is performed, as in the downlink transmission, through the DCF scheme. Both basic access and handshaking (RTS/CTS) mechanisms can be used.

\end{itemize}

\paragraph{Sub-Slotting Mechanisms}
To ensure maximum channel occupancy and energy savings, TGah also sets the option to split uplink and downlink restricted access windows into several time slots. Thus, each slot contains only a few stations with data to receive/send. When there is only one station assigned per slot, the mechanism becomes a regular TDMA.

Because of the sub-slotting, stations belonging to the same TIM group could also be distributed over different sub-slots. Therefore, they would save even more energy by competing for the channel with fewer stations and extending the time spent in the sleeping mode \cite{EW_subm}. 

\subsection{Long Sleeping Periods}

All stations regulated by the \textit{TIM and Page Segmentation} scheme are forced to listen to every DTIM beacon, even when they have long intervals during which they are unlikely to have data to send or receive. This is why IEEE 802.11ah also offers, to any kind of station, the possibility of fixing very long doze times (up to years) during the initial handshake with the AP. However, the clock drift produced by such long doze times may become a major synchronization problem. The reason is that the longer a station has been asleep, the further in advance it should wake up to avoid possible synchronization lags with the network.

\subsection{Support for Small Data Transmission}

Small data transmissions usually generate high overheads and low performance in wireless sensor networks. To reduce the impact of these overheads, TGah proposes the use of three new enhancements. First, while IEEE 802.11 uses a 28-byte MAC header, IEEE 802.11ah shortens it to 18 bytes by means of using association identifiers instead of regular MAC addresses \cite{zhou2013}. 

Second, TGah has defined several null data packet frames, that consist of a single PHY header to shorten current IEEE 802.11 signaling frames, such as ACKs, block ACKs, CTSs and PS-Polls. 

Finally, a speed frame exchange \cite{Draft802.11ah} mechanism has been included to signal the successful reception of frames by transmitting a data frame instead of a traditional ACK.


\section{Performance Assessment}
\label{proposal}

We used a one-hop fully connected network in all the IEEE 802.11ah simulations where we evaluated two outdoor (agriculture and animal monitoring) and two indoor (smart metering and industrial automation) scenarios.



In all four scenarios, stations were uniformly distributed within the AP coverage. The speed rate for each transmission depends on the relative distance between the AP and the station, and it is obtained from the propagation model defined by \cite{Hazmi2012}, for both the indoor and outdoor scenarios. The relationship between distance and data rate is shown in Figures \ref{fig:out_model} and \ref{fig:ind_model}, respectively. In all simulations we fixed a packet error rate of 10\%.

Each station was capable of receiving and transmitting only one data packet per DTIM interval. These intervals were split into several TIM intervals, whose downlink and uplink restricted access window segments were occupied only by TIM-type stations. RTS/CTS mechanism is only implemented in the uplink. The restricted access window segment size ($\psi \in \lbrace \DL,\UL\rbrace$) is proportional to the traffic fraction $(\beta_{\psi})$ delivered in each. We also assume a multicast restricted access window with a capacity of one data packet.

The simulations were performed in MATLAB, and the main parameters are presented in Table \ref{fig:sim_param}.

\renewcommand{\figurename}{Table}%
\setcounter{figure}{1}
\begin{figure}
\begin{center}
\subfigure[List of PHY layer simulation parameters 
\label{fig:model_phy}]
{\includegraphics[height=6.3cm]{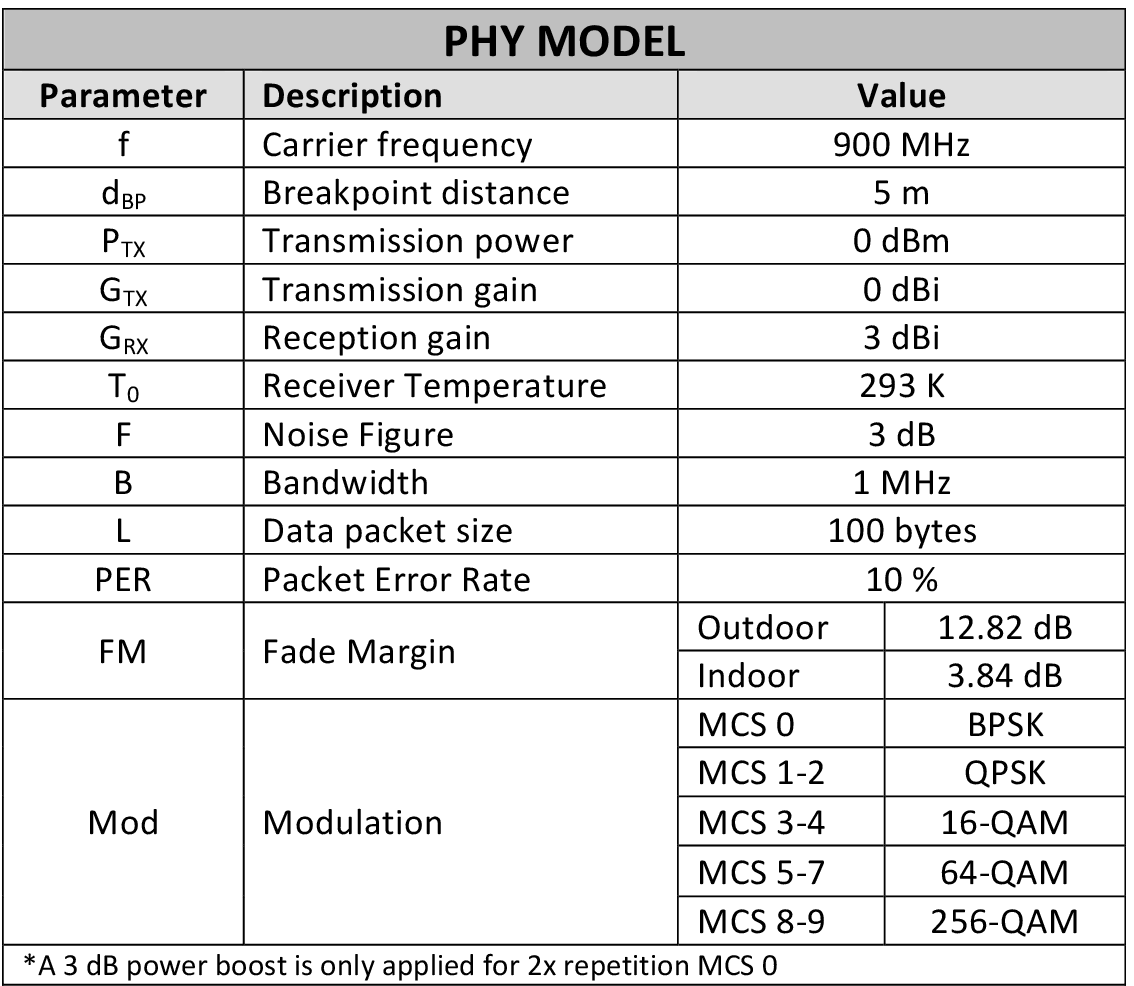}}
\subfigure[List of MAC layer simulation parameters  
\label{fig:model_mac}]
{\includegraphics[height=6.3cm]{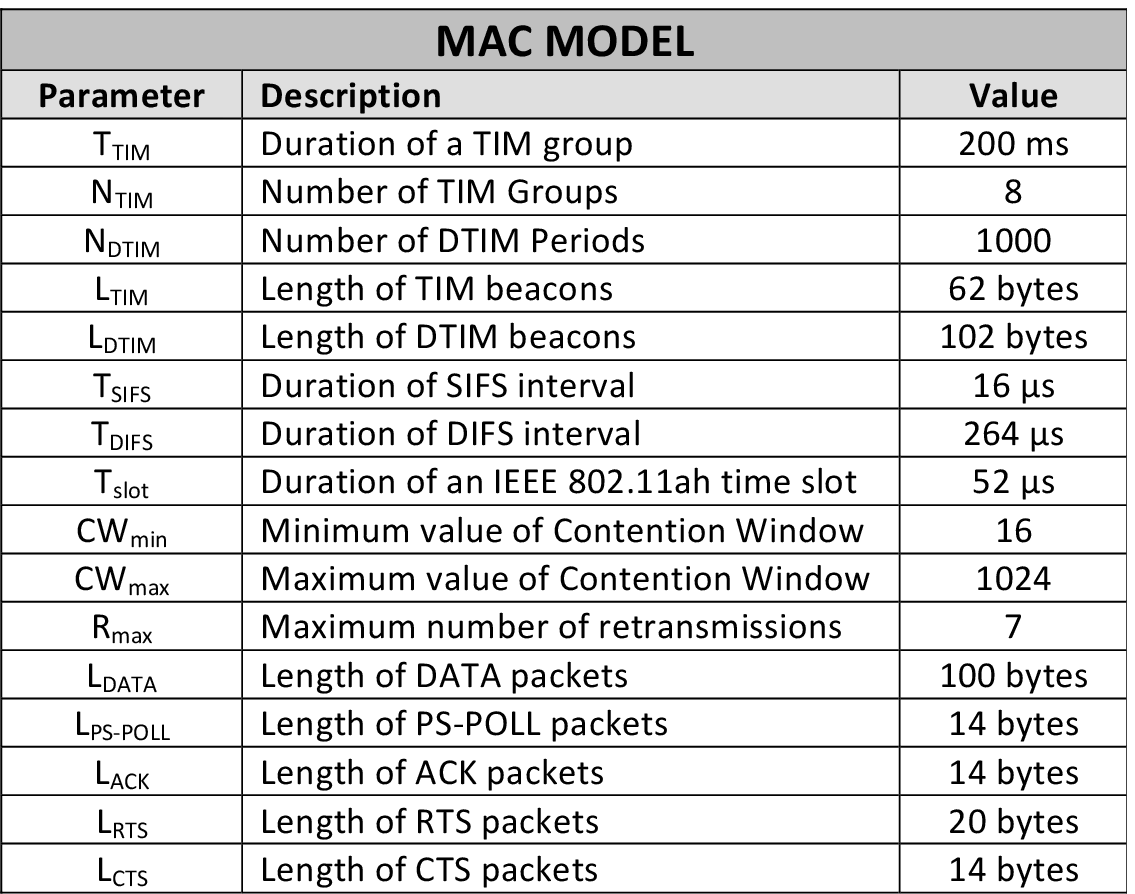}}
\end{center}
\caption{Main simulation parameters for PHY and MAC layers}
\label{fig:sim_param}
\end{figure}

\renewcommand{\figurename}{Figure}%

Four performance metrics were analyzed in each scenario:

\begin{itemize}
 \item Packet delivery ratio ($\text{PDR}_{\psi}$) represents the percentage of packets that successfully reached their destination versus the number of packets generated.

\item Packet delivery delay ($\text{PDD}_{\psi}$) measures the time to complete a successful transmission in the number of DTIM beacons.
 
\item The channel occupancy ($\eta_{\psi}$) is the ratio between the number of packets delivered successfully and the theoretical capacity.

\item Energy consumption and battery duration depends on the time spent in receiving, transmitting, idle or sleeping modes. Each mode is described as follows:

\begin{itemize}
\item \textbf{Receiving}: Stations in power saving mode that have not entered into a long sleeping period must listen to all the DTIM beacons. If a station is signaled in a DTIM beacon with downlink data, or it has data to transmit, it will also listen to its corresponding TIM beacon. An station receiving a data packet, a CTS or an ACK is in the receiving state as well. Overhearing packets addressed to other stations also keeps stations in receiving mode.
	\item \textbf{Transmitting}: Stations sending frames in both downlink (PS-Poll, ACK) and uplink (RTS, DATA) channels.
	\item \textbf{Idle}: The backoff periods and any inter-frame spaces.
	\item \textbf{Sleeping}: Stations whose radio module is switched off. 
\end{itemize}

\end{itemize}

\subsection{Scenarios}

As stated above, four common application scenarios have been used to evaluate the performance of IEEE 802.11ah WLANs: agriculture monitoring, smart metering, industrial automation and animal monitoring. The parameters used are summarized in Figure \ref{fig:sce_param}. All scenarios shared the same downlink inter-arrival time, that is, the time between two consecutive packets, fixed at a value of 4 minutes \cite{scoop} .

\setcounter{figure}{1}
\begin{figure}
\begin{center}
\subfigure[Outdoor propagation model   \label{fig:out_model}]
{\includegraphics[width=5cm]{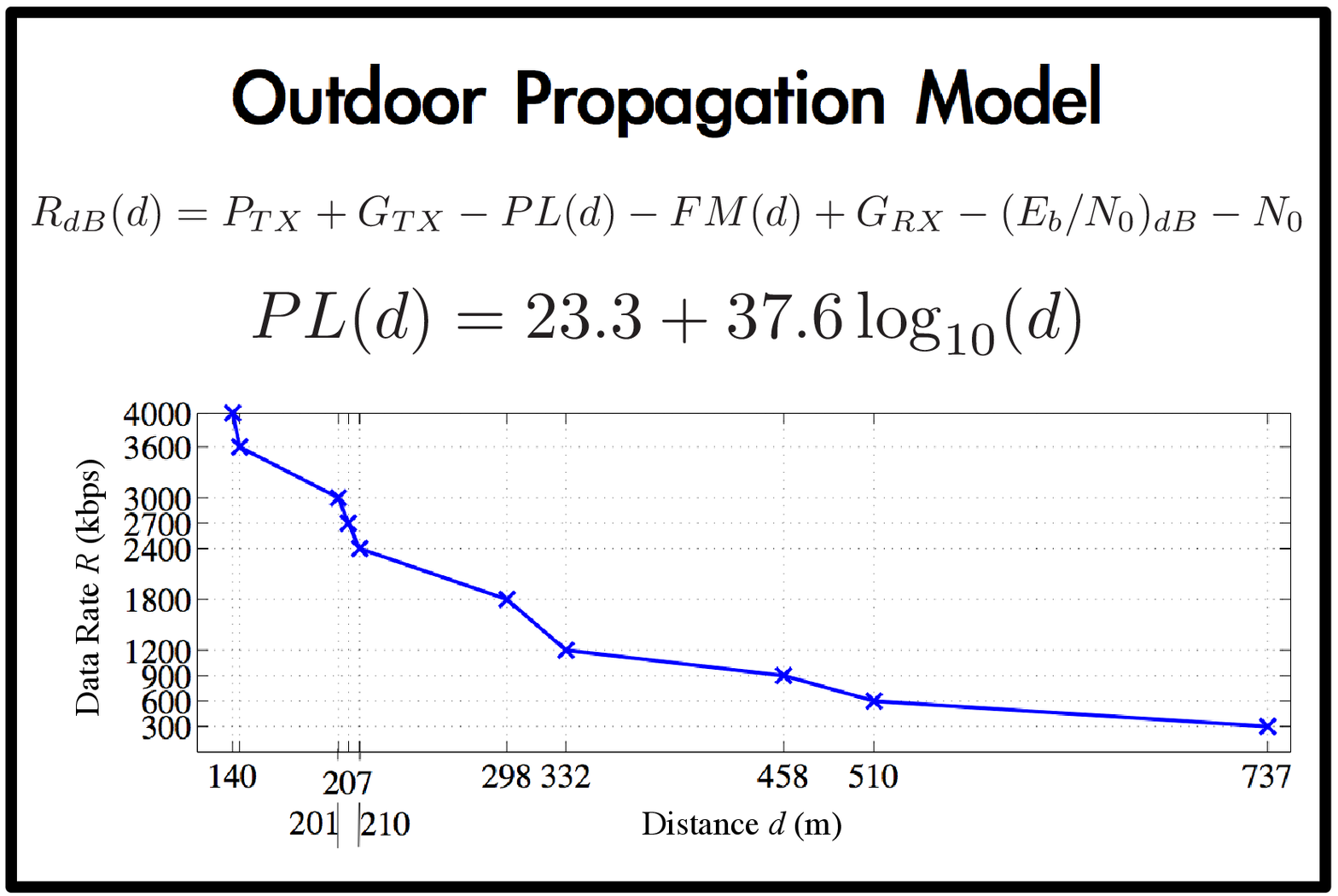}}
\subfigure[Outdoor scenarios: Agriculture and animal monitoring   
\label{fig:out_sce}]
{\includegraphics[width=10cm]{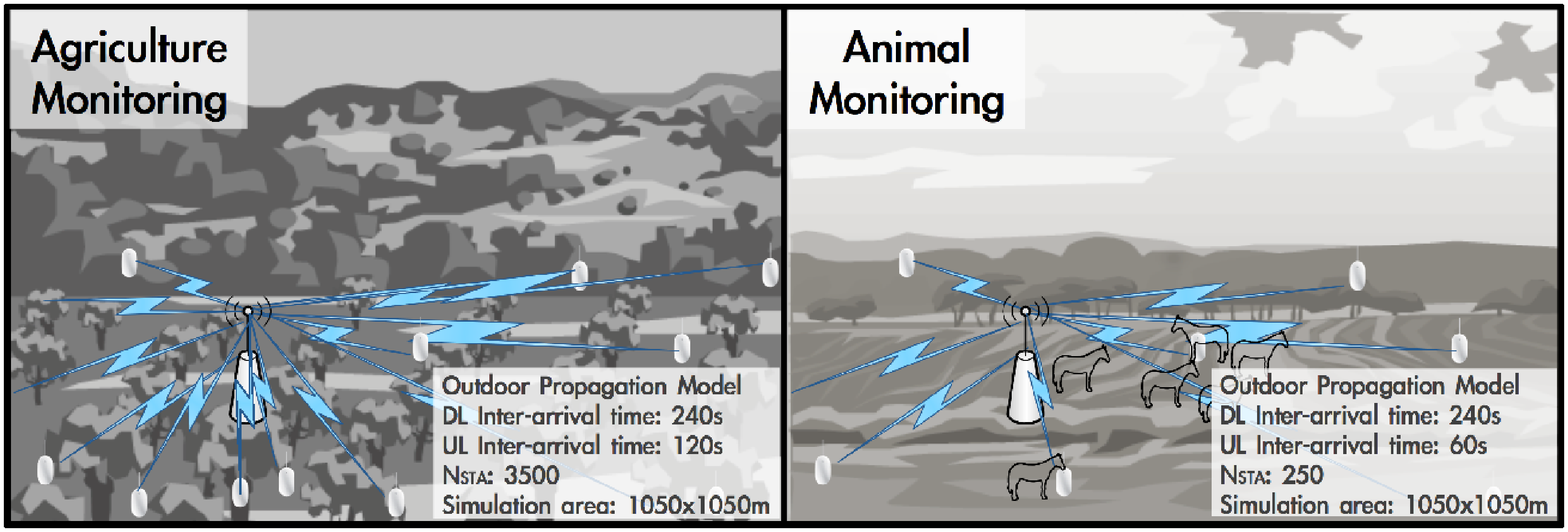}}
\subfigure[Indoor propagation model \label{fig:ind_model}]
{\includegraphics[width=5cm]{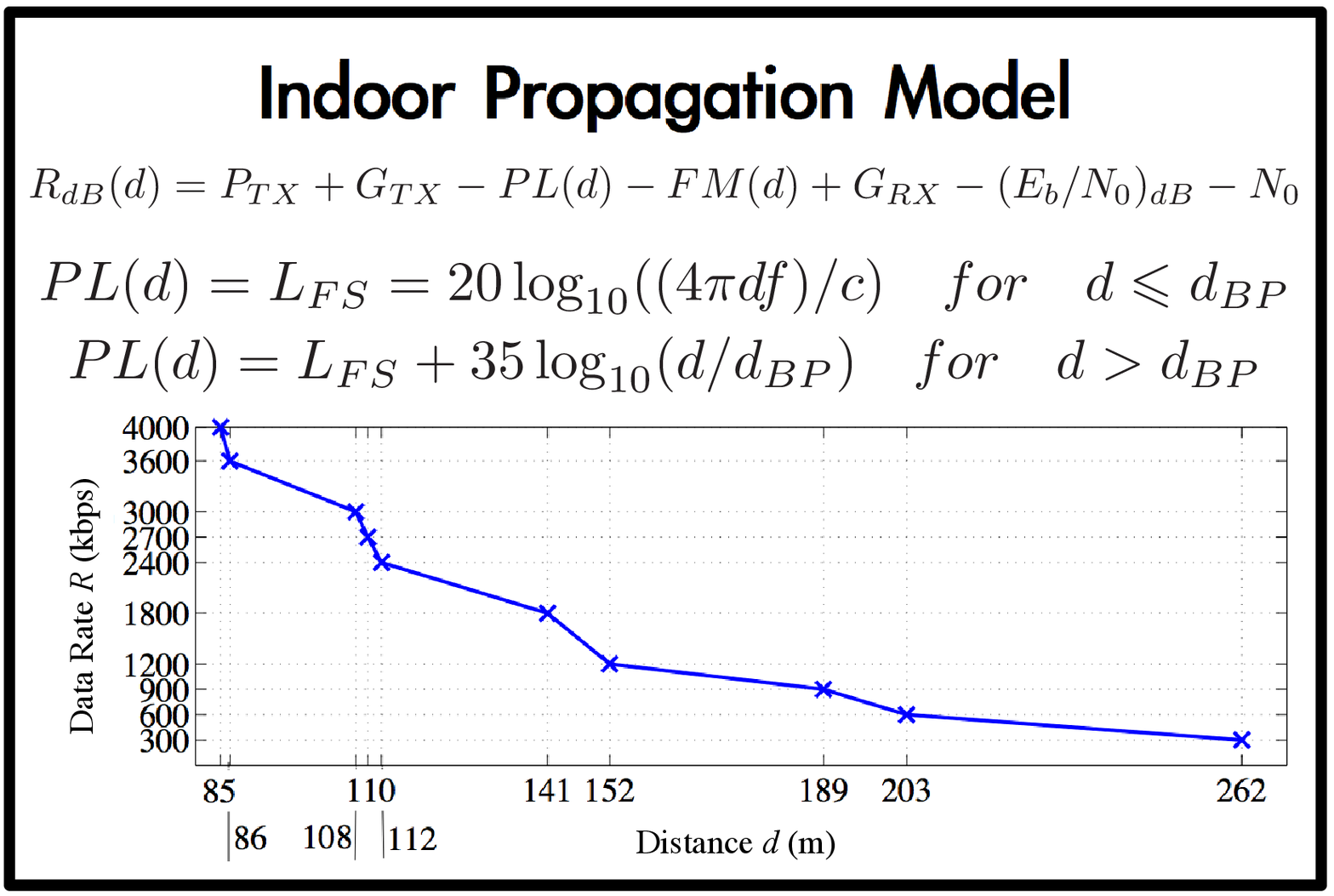}}
\subfigure[Indoor scenarios: Industrial automation and smart metering  
\label{fig:ind_sce}]
{\includegraphics[width=10cm]{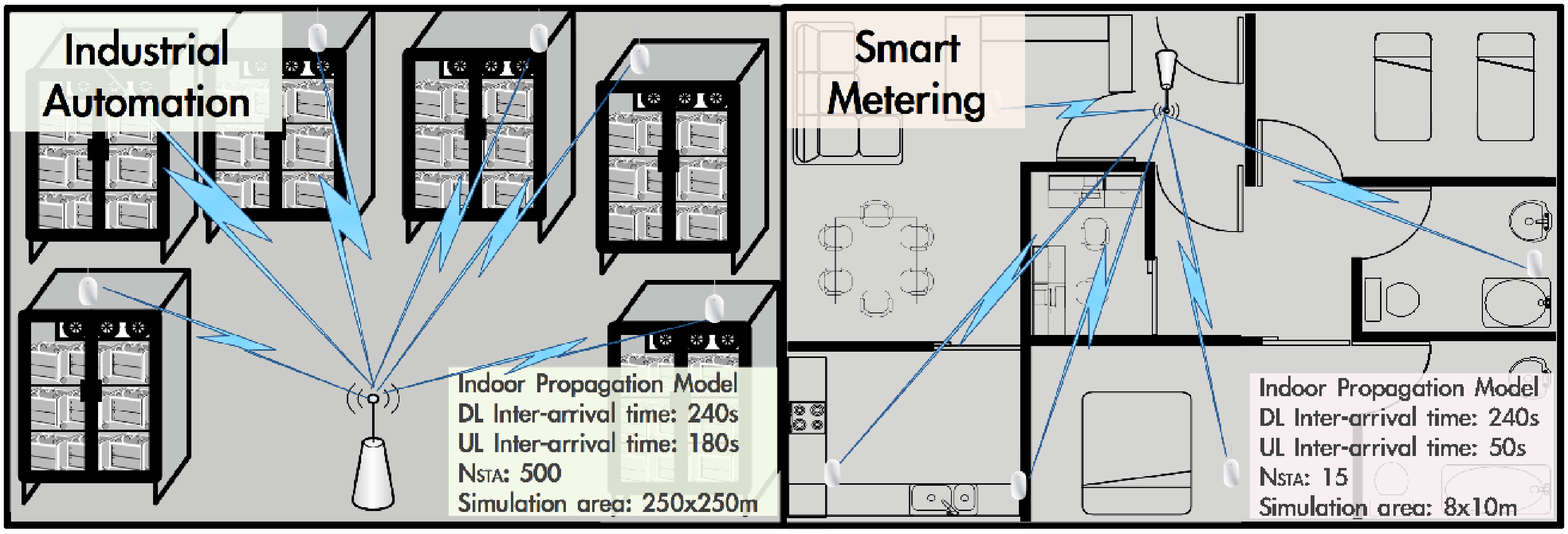}}
\caption{Application scenarios and the simulation parameters used in each} \label{fig:sce_param}
\end{center}
\end{figure}

\subsubsection{Agriculture monitoring}

In \cite{scenario1}, a WSN was used to control the environmental humidity of an agricultural field to activate the irrigation system. In this scenario, 3,500 sensor nodes were deployed, transmitting a message every 120 seconds. 
 
\subsubsection{Smart metering}

In \cite{scenario3}, indoor sensors reported home measurements of the electricity consumption from several electrical appliances. This scenario counts the presence of $15$ sensors, each of them transmitting data every 50 seconds.

\subsubsection{Industrial automation}

In this third scenario, we considered the application of temperature control in refrigerating chambers \cite{scenario2}. The goal is to monitor the temperature of the refrigerators to ensure stable conservation conditions for fruit and vegetables. To achieve the goal, a sensor network is deployed inside the chambers to report measurements every 180 seconds. To control the temperature conditions and detect breaks in the cooling chain inside an industrial warehouse, hundreds of sensor stations are required over the area. In this simulation scenario, we used 500 stations.

\subsubsection{Animal monitoring}

Currently, the control of animals that inhabit protected natural areas is performed manually, which is costly and may provoke stress to the animals when they are being manipulated. Thus, sensor networks are a useful solution as proposed in \cite{scenario4}. In this scenario, sensor measurements are sent every 60 seconds from a network with 250 devices.

\subsection{Channel Occupancy}

The results of channel occupancy,($\eta_{\psi}$), are shown in Table \ref{table:occupancy}, where we observe a low utilization of the available resources in all four scenarios. 

As expected, agriculture and animal monitoring scenarios show a higher channel occupancy than the others. Both are outdoors and have to cover large areas, demanding distant stations to transmit at low data rates. Hence, their transmissions are longer and, as a result, the channel occupancy is higher. Moreover, the larger number of stations (3,500) in the agriculture monitoring scenario increased the use of the channel resources compared to the case of the animal monitoring application with only 250 devices.

The channel occupancy for indoor scenarios ($\eta_{\psi}$) is far below 1\%. With those results, we reflect that the IEEE 802.11ah amendment has the ability to efficiently manage a large number of stations with a single AP in a wide set of applications. A more detailed study of the IEEE 802.11ah network capacity can be found in \cite{adame2013}.

\renewcommand\thetable{3}
\setcounter{table}{2} 
\begin{table}[h!]
\begin{center}
\begin{tabular}{|l|c|c|c|c|c|c|}
\hline
 & \multicolumn{2}{|c|}{\textbf{$\eta_{\psi}$} (\%)} & \multicolumn{2}{|c|}{\textbf{$PDR_{\psi}$} (\%)} & \multicolumn{2}{|c|}{\textbf{$PDD_{\psi}$} (s)}\\
\hline 
\textbf{Scenario} & \textbf{Downlink} & \textbf{Uplink} & \textbf{Downlink} & \textbf{Uplink} & \textbf{Downlink} & \textbf{Uplink}\\
\hline
Agriculture monitoring & 8.79 & 8.92 & 100 & 100 & 0.22 & 0.23 \\
\hline
Smart metering & 0.022 & 0.024 & 100 & 100 & 0.2 & 0.2\\
\hline
Industrial automation & 0.46 & 0.57 & 100 & 99.98 & 0.28 & 0.29\\
\hline
Animal monitoring & 1.1 & 1.3 & 100 & 99.94 & 0.23 & 0.26\\
\hline
\end{tabular}
\vspace{-0.3cm}
\end{center}
\caption{Simulation results for the scenarios considered}
\label{table:occupancy}
\end{table}

\subsection{Packet Delivery Ratio and Packet Delivery Delay}

In all the analyzed scenarios, the results achieved for the packet delivery ratio are 100\% for downlink traffic and close to 100\% for uplink traffic. As expected, because of the low values of channel occupancy ($\eta_{\psi}$), the network has been able to transmit nearly all generated traffic, despite some minor retransmissions due to packet errors or collisions.  

\subsection{Energy Consumption and Battery Estimation}
\renewcommand\thefigure{3} 
\begin{figure}[p]
\begin{center}
\subfigure[Percentage of time spent in Scenario 1: Agriculture monitoring \label{fig:agriculture}]
{\includegraphics[width=0.5\textwidth]{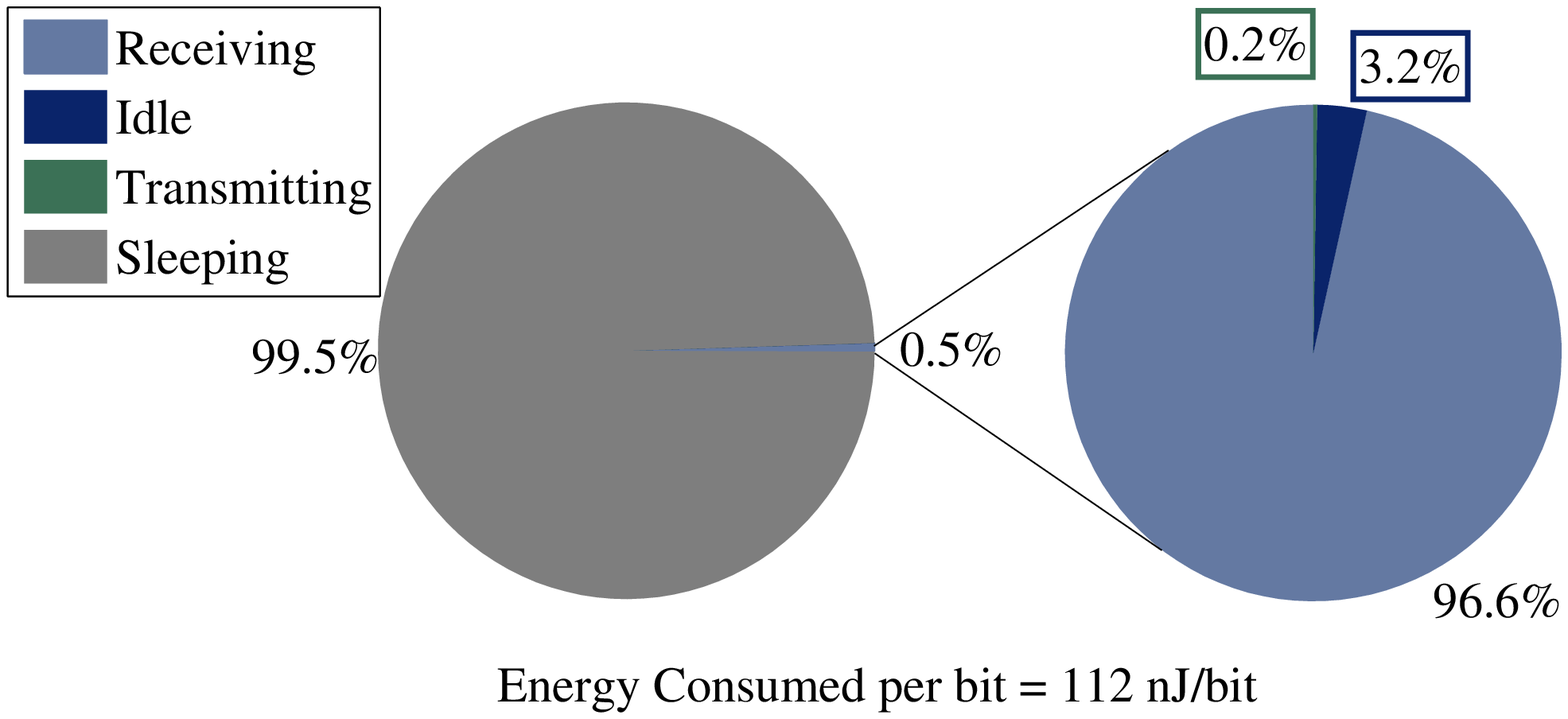}}
\subfigure[Percentage of time spent in Scenario 2: Smart metering \label{fig:domotica}]
{\includegraphics[width=0.5\textwidth]{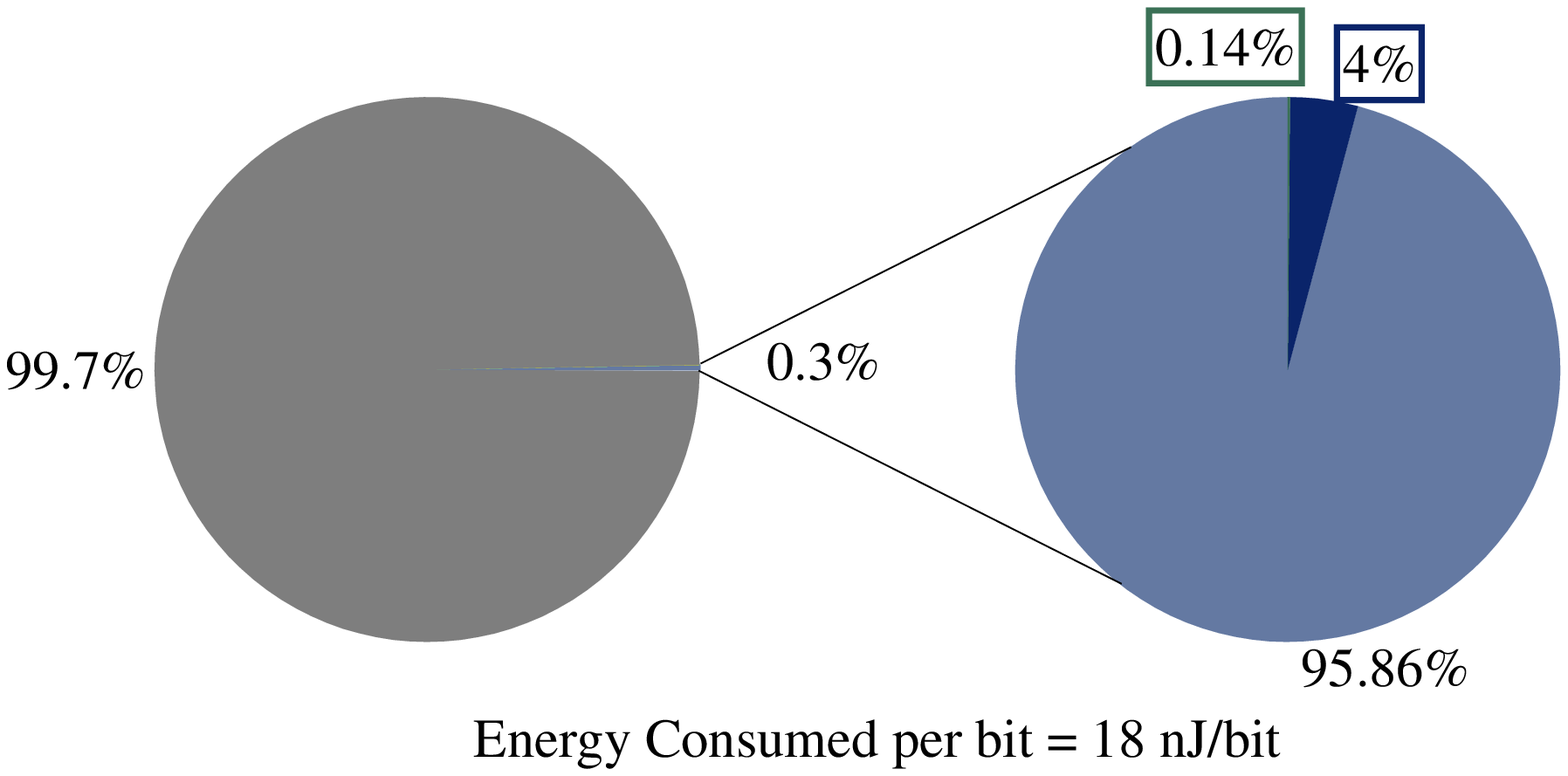}}
\subfigure[Percentage of time spent in Scenario 3: Industrial automation \label{fig:industrial}]
{\includegraphics[width=0.5\textwidth]{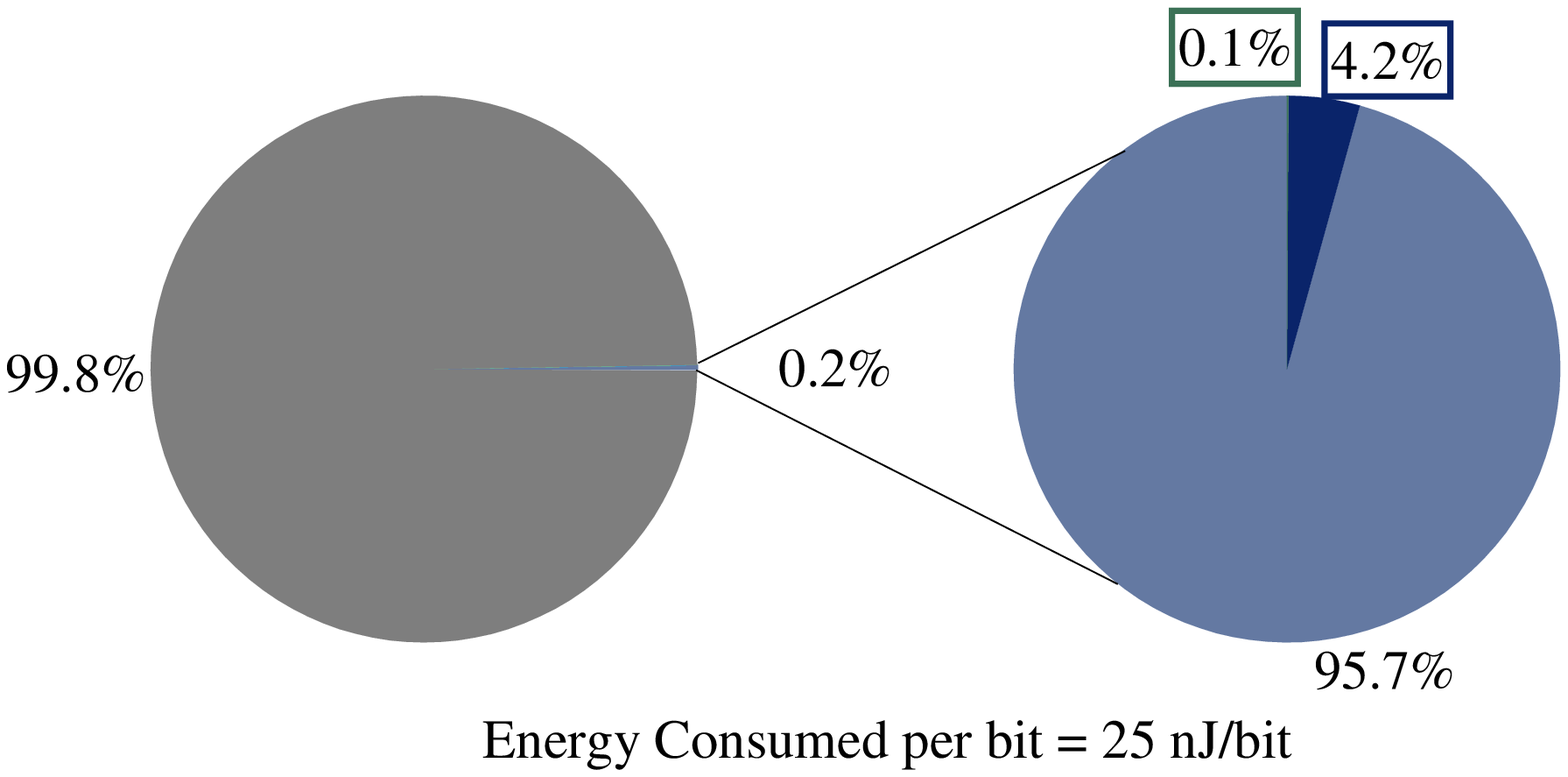}}
\subfigure[Percentage of time spent in Scenario 4: Animal monitoring \label{fig:parking}]
{\includegraphics[width=0.5\textwidth]{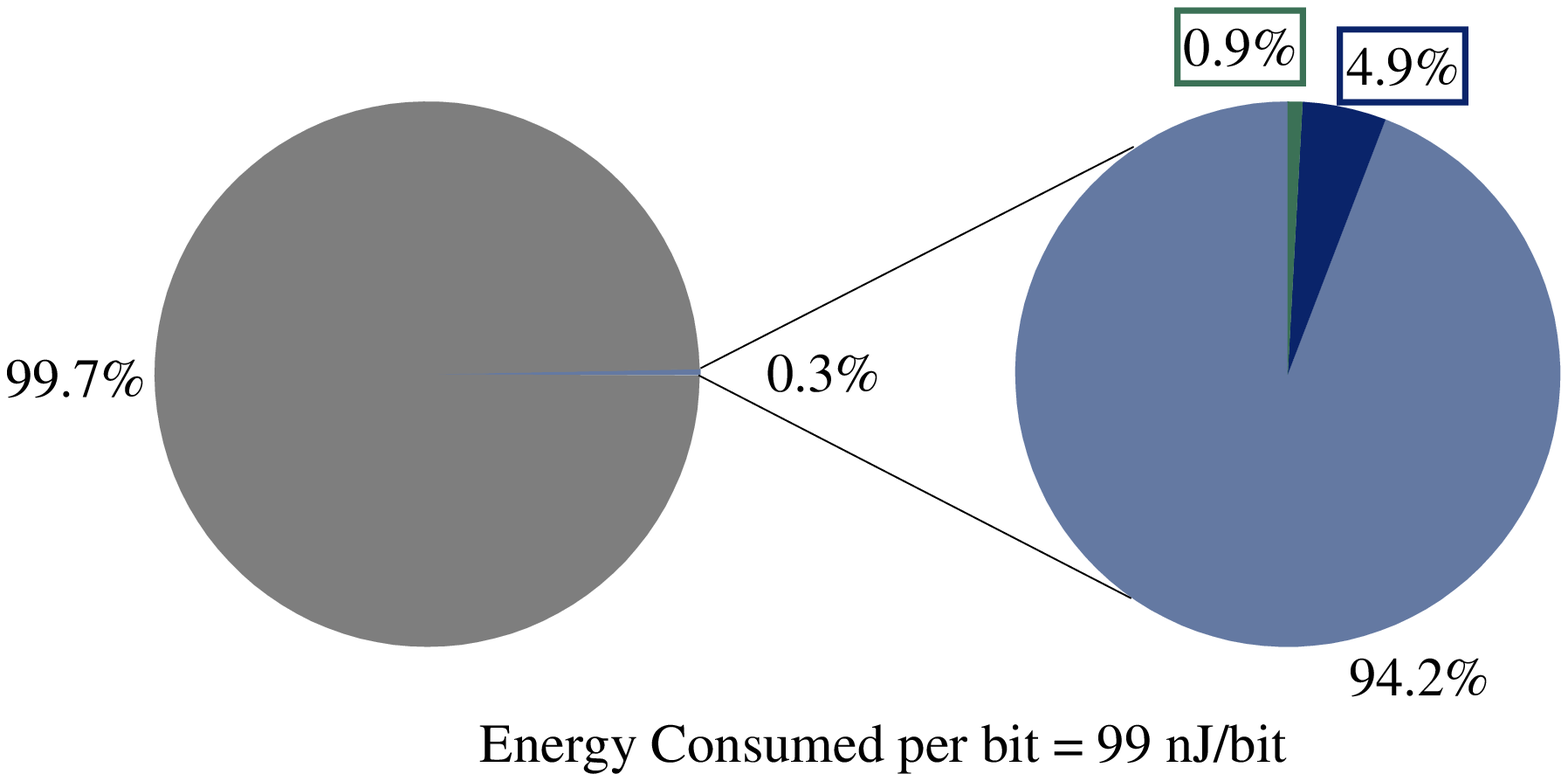}}
\subfigure[Battery lifetime\label{fig:Battery}]
{\includegraphics[width=0.65\textwidth]{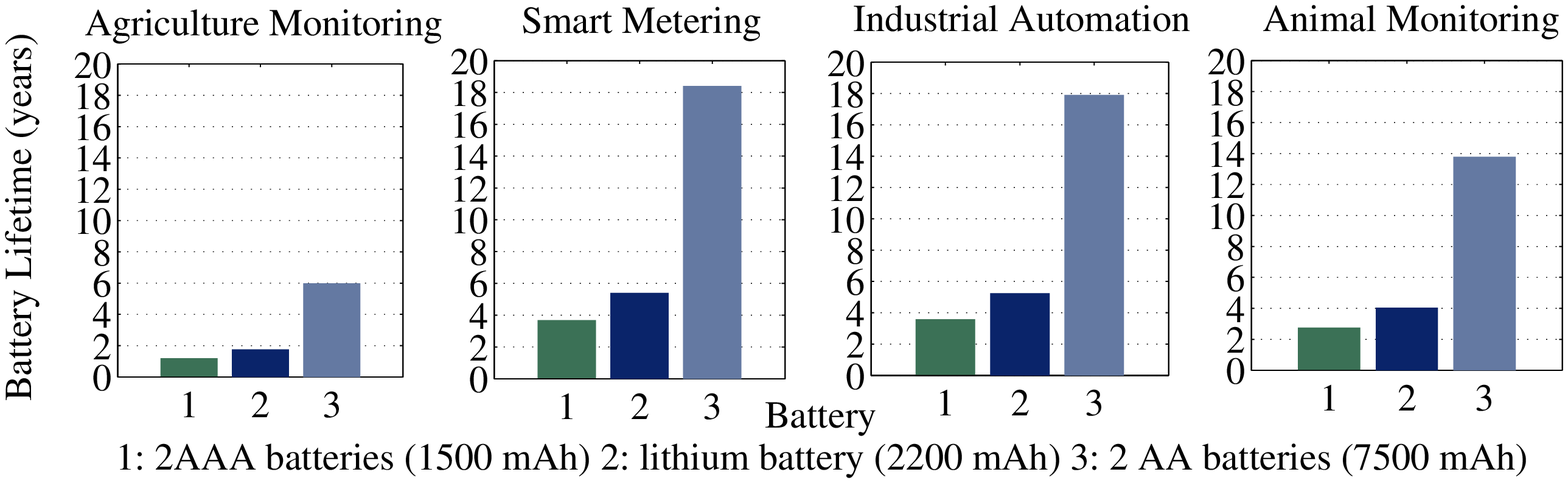}}
\caption{Time spent in each state, energy consumed per bit and battery lifetime of a CC1100 transceiver performing IEEE 802.11ah in the proposed scenarios} \label{fig:energy}
\end{center}
\end{figure}

The results regarding the time spent in each state and the energy consumed per bit are shown in Figure \ref{fig:energy}. To obtain these results, the transition time between two different states is assumed to be negligible and therefore is assumed to consume no energy. In any case, because of the low periodicity of those transitions, its impact on the overall energy consumption is expected to be minor.

It is worth noting that a node, on average, will remain in the sleeping state most of the time (close to 99\%) in all scenarios. Hence, the energy consumption can be assumed to be very low. The highest consumption is obtained, as expected, in the agricultural monitoring scenario because of the assumed data rate and the presence of a large number of stations.

In the simulations, we modeled different types of batteries (Figure \ref{fig:Battery}). The maximum battery duration, 18 years, is achieved in the smart metering scenario. The lowest duration, approximately 6 years, occurs in the agricultural scenario. It should be noted that, in each scenario, the battery lifetime is calculated for the worst case, which is the node with the highest energy consumption.

\section{Conclusions} \label{conclusions}

The limitations that currently do not allow Wi-Fi to play an important role in M2M communications are solved with the adoption of the new IEEE 802.11ah amendment. Its new energy-saving mechanisms ensure an efficient use of the limited energy resources available in sensor nodes. Moreover, its operation in a sub-1GHz band achieves larger coverage areas than  the original IEEE 802.11. The number of simultaneously operable stations has also been increased, up to 8,191, where all the devices can be managed by a single AP using a new hierarchical organization.

In this paper, we have evaluated the feasibility of the new IEEE 802.11ah amendment. Because of the infrequent data exchange in M2M applications, a large number of stations can share a single IEEE 802.11ah AP, as long as their activity periods are properly distributed over time. In addition to the communication needs, energy efficiency becomes a critical issue in applications based on battery-powered nodes. On average, the results obtained show that stations remained in sleeping mode more than $99$\% of the time, demonstrating a higher energy efficiency of IEEE 802.11ah.

In the future, due to the heterogeneous requirements of M2M applications, the definition of new QoS differentiation mechanisms will be needed to allow the coexistence of different M2M applications within the same AP coverage. In addition, there are still open challenges regarding the performance of Non-TIM and unscheduled stations, as well as their integration with TIM stations in a single WLAN. Moreover, there is also a lack of comparative performance studies between the different existing and future technologies for M2M communications. Such comparative studies would provide a set of guidelines to select the best technology to use in a given scenario.

\section*{Acknowledgements}
This work was partially supported by the Spanish government through the projects TEC2012-32354 and IPT-2012-1028-120000.

\bibliographystyle{unsrt}
\bibliography{Bib}

\end{document}